\documentclass[prb, preprint, unsortedaddress, showpacs, nofootinbib, eqsecnum]{revtex4}
\usepackage{amsfonts, amsbsy, amssymb, amsmath, graphicx,  float, color} 
\usepackage{subfigure}

\newcommand{\rmd}{{\rm d}}
\newcommand{\deriv}[2]{{\frac{\rmd {#1}}{\rmd {#2}}}}

\newcommand{\calH}{\mathcal{H}}

\newcommand{\bu}{\boldsymbol{g}}
\newcommand{\bv}{\boldsymbol{v}}

\newcommand{\br}{\boldsymbol{r}}

\begin{document}


\title{Nonstatistical dynamics on potentials exhibiting 
reaction path bifurcations and valley-ridge inflection points}



\author{Peter Collins}
\affiliation{School of Mathematics  \\
University of Bristol\\Bristol BS8 1TW\\United Kingdom}

\author{Barry K. Carpenter}
\email[]{CarpenterB1@cardiff.ac.uk}
\affiliation{School of Chemistry\\
Cardiff University\\
Cardiff\\
CF10 3AT\\
United Kingdom
}

\author{Gregory S. Ezra}
\email[]{gse1@cornell.edu}
\affiliation{Department of Chemistry and Chemical Biology\\
Baker Laboratory\\
Cornell University\\
Ithaca, NY 14853\\USA}

\author{Stephen Wiggins}
\email[]{stephen.wiggins@mac.com}
\affiliation{School of Mathematics \\
University of Bristol\\Bristol BS8 1TW\\United Kingdom}


\date{\today}

\begin{abstract}

We study reaction dynamics on a model potential energy surface 
exhibiting post-transition state bifurcation in the vicinity of
a valley ridge inflection point.
We compute fractional yields of 
products reached after the VRI region is traversed, both with and
without dissipation.
It is found that 
apparently minor variations in the potential lead to significant changes in the reaction
dynamics.  Moreover, when dissipative effects are incorporated, 
the product ratio depends in a complicated and highly non-monotonic fashion
on the dissipation parameter.
Dynamics in the vicinity of the VRI point itself play essentially no role in
determining the product ratio, except in the highly dissipative regime.

\end{abstract}


\maketitle


\section{Introduction}

Much recent experimental and theoretical work 
has focussed on recognizing and understanding the manifestations of nonstatistical
dynamics in thermal reactions of organic molecules 
(for reviews, see refs \onlinecite{Carpenter92,Carpenter04,Carpenter05,Bachrach07,Birney10,Hiroshi10,Rehbein11};
see also the representative refs 
\onlinecite{Ussing06,Thomas08,Ess08,Wang09,Oyola09,Hong09,Glowacki09,Katori10,Siebert11,Goldman11,Quijano11,Yamamoto11}). 
Such research has convincingly demonstrated that, for an ever-growing 
number of cases, standard transition state theory (TST) and RRKM approaches 
\cite{Wigner38,Bunker66,Robinson72,Forst73,Baer96,Truhlar96,Forst03,Henriksen08}
for prediction of rates, 
product ratios, stereospecificity and isotope effects can fail completely.  
This work is changing the basic textbook paradigms 
of physical organic chemistry (cf.\ ref.\ \onlinecite{Bachrach07}, Ch.\ 7).

While absolute rates are usually controllable with changes of temperature, 
relative rates (i.e., \emph{selectivity}) often are not \cite{Carpenter84}.  
Hence, understanding the factors that control selectivity is of essential importance 
for synthesis, 
especially if existing models used for analyzing the problem are incomplete or inapplicable.

A fundamental dynamical assumption underlying conventional statistical theories of
reaction rates and selectivities 
is the existence of intramolecular vibrational energy
redistribution (IVR) that is rapid compared to the rate of reaction/isomer\-ization
\cite{Tabor81,Rice81,Brumer81,Noid81,Brumer88,Uzer91}.
Such rapid IVR leads to a `loss of memory' of particular initial conditions \cite{Brass93}.
Standard computations based on features (usually critical points, such as
minima and saddle points) 
of the potential energy surface 
(PES) then provide predictions for relative rates associated with competing reactive channels,
temperature dependence of reaction rates and branching ratios, etc
\cite{Wigner38,Bunker66,Robinson72,Forst73,Carpenter84,Baer96,Truhlar96,Forst03,Wales03,Henriksen08}.
Nonstatistical effects can arise from a number of factors, which are certainly
not mutually exclusive (see, for example, refs 
\onlinecite{Bunker62,Bunker64,Hase76,Hase98,Grebenshchikov03,Bach06,Lourderaj09},
also \onlinecite{Schofield94,Schofield95,Schofield95a,Leitner96,Leitner97a,Hase94,Gruebele04,Leitner06,Leitner08,Leitner11}). 
The essential underlying 
reason is the `failure of ergodicity', a property 
which is notoriously difficult either to predict
or diagnose.  Branching ratios and/or stereochemistries 
significantly different from statistical predictions can result from symmetry
breaking induced by dynamics \cite{Carpenter05}.

The range of thermal organic reactions now believed to manifest some kind of nonstatistical behavior
is extraordinarily diverse (see, for example, refs \onlinecite{Carpenter05,Rehbein11}, and refs therein).
              A general characteristic shared by  the systems for which the standard statistical theories
	      fail is that the associated PES corresponds 
	      poorly if at all to the standard textbook picture of a 
	      1D reaction coordinate passing over high barriers connecting
	      deep wells (intermediates or reactants/products, cf.\ Fig.\ \ref{fig:bkc_1}a)
	      \cite{Carpenter04,Carpenter05,Rehbein11}.  More specifically, 
	      the reaction coordinate (understood in the broadest sense 
	      \cite{Heidrich95,Kraka11}) is inherently
	      \emph{multidimensional}, as are  corresponding relevant phase space
	      structures: there may exist extremely 
	      flat, plateau regions on the PES, with a number of exit channels characterised by
	      low barriers \cite{Carpenter04,Carpenter05}, 
	      or the PES may exhibit \emph{bifurcations} of the reaction path 
	      \cite{Valtazanos86,Quapp98,Quapp04,Lasorne05,Sugny06}
	      in the vicinity of so-called valley-ridge inflection (VRI) points 
	      \cite{Thomas08,Ess08,Wang09,Siebert11} (see Fig.\ \ref{fig:bkc_1}b).  In the case of systems 
	      having dynamically relevant VRI points on/near the reaction path, outstanding fundamental questions
	      remain concerning the effectiveness of approaches such as variational TST 
	      \cite{Keck67,Hase83a,Truhlar84} or modified statistical theories \cite{Zheng09}
	      as opposed to full-scale trajectory simulations of the reaction dynamics
\cite{Gonzalez04,RamirezAnguita11}.

Other systems for which the standard 1D reaction coordinate picture is not valid
include the growing class of so-called non-MEP (minimum energy path) reactions
\cite{Mann02,Sun02,Debbert02,Ammal03,Carpenter04,Lopez07,Lourderaj08,Lourderaj09} and ``roaming'' mechanisms
\cite{Townsend04,Bowman06,Shepler07,Shepler08,Suits08,Heazlewood08,Bowman11,Bowman11a}; 
the dynamics of these reactions is not mediated by
a single conventional transition state associated with an index 1 saddle.

The work just described highlights the basic importance of
\emph{momentum} for the outcome of a chemical reaction \cite{Carpenter98,Carpenter05};
that is, a \emph{phase space} \cite{MacKay87}
approach to reaction dynamics is needed \cite{Waalkens08}, as opposed to
a view wedded solely to the topography of the PES \cite{Mezey87,Wales03}.  
For example, a symmetric
PES with two symmetry-related reaction channels 
can give rise to asymmetric product distributions if nonsymmetric initial momentum 
distributions are created under experimental conditions \cite{Carpenter05}.

There have 
been significant recent theoretical and computational advances in the 
application of dynamical systems theory  \cite{MacKay87,Lichtenberg92,Wiggins92,Arnold06} 
to study reaction dynamics and phase space structure in multimode models of 
molecular systems, and to probe the
dynamical origins of nonstatistical behavior 
\cite{Wiggins90,wwju,ujpyw,WaalkensBurbanksWiggins04,WaalkensWiggins04,WaalkensBurbanksWigginsb04,
WaalkensBurbanksWiggins05,WaalkensBurbanksWiggins05c,SchubertWaalkensWiggins06,Waalkens08,Ezra09,Ezra09a,Collins11}
(see also refs
\onlinecite{Komatsuzaki00,Komatsuzaki02,Toda02,Komatsuzaki05,Wiesenfeld03,Wiesenfeld04,Wiesenfeld04a,Toda05,Gabern05,Gabern06,Shojiguchi08}).
A phase space approach is essential to obtain a rigorous 
dynamical definition of the TS in multimode systems,
this being the Normally Hyperbolic Invariant Manifold (NHIM) \cite{Waalkens08}.
The NHIM generalizes the concept of the
periodic orbit dividing surface (PODS) \cite{Pollak78,Pechukas81,Pechukas82}
to $N \geq 3$ mode systems.
A recent reappraisal of the \emph{gap time} formalism for 
unimolecular rates \cite{Ezra09a} has led to novel 
diagnostics for nonstatistical behavior (`nonexponential decay') in
isomerization processes, leading to a necessary condition for ergodicity.

In the present paper we study dynamics on a model potential energy surface 
(PES) exhibiting post-transition state bifurcation in the vicinity of
a valley ridge inflection point (cf.\ Fig.\ \ref{fig:bkc_1}b).  
A computed normal form (NF) is used to sample the dividing surface (DS) at fixed total energy 
at the `incoming' TS located at a point of high potential energy.  
(For previous discussion of sampling using normal forms, see ref.\ 
\onlinecite{Collins11}.)
Bundles of trajectories so defined
are then followed into the region of the PES where bifurcation of the reaction path occurs, 
and the subsequent dynamics studied. 
A key goal here is to obtain a dynamical understanding  of the 
computed \emph{branching ratio} for products reached after the VRI region is traversed. 
By changing parameters in the model potential, it is for example possible to
alter the location but not the energy of one of the product minima, while keeping 
the energies and locations of other critical points unchanged.  We find that 
apparently minor variations in the potential can lead to significant changes in the reaction
dynamics.

The work described below
leads to the following picture of the dynamical 
origin of the selectivity: for the model studied,  
the dynamics proceeds on at least 2 timescales.  First, on short times, the
bundle of trajectories `reflects' off a hard wall that is opposite the
high energy TS through which it enters the reaction zone.  
After this collision, a highly non-statistical and time-dependent 
population ratio of products is established,
whose value depends on the direction in which trajectories are reflected by
the (asymmetric) potential wall. These initial
nonstatistical populations then relax on a somewhat longer timescale to 
yield the observed product ratio.  
During this phase of the reaction, there is 
the possibility of competition between IVR and other mechanisms 
for removing vibrational energy from active degees of freedom (cf., for example, ref.\ \onlinecite{Quijano11}).
We explore one aspect of this phase of the  dynamics by introducing dissipation
into the model.
Major conclusions are that, for this model at least, 
(i) the dynamics in the vicinity of the VRI point plays essentially no role in
determining the product ratio, except in the highly dissipative regime, 
and (ii) the product ratio is a highly nonmonotonic
function of the dissipation strength.

The structure of this paper is as follows: in Sec.\ \ref{sec:pes} we introduce the model 
potential energy function to be studied.  We compute the 
IRC path \cite{Fukui70} connecting the TS to one or the other product minima, discuss the 
location of the VRI point \cite{Valtazanos86,Quapp98,Quapp04}, and compute 
Newton trajectories \cite{Quapp98a,Hirsch04} and gradient extremals 
\cite{Rowe82,Hoffman86}.
In Section \ref{sec:traj} we formulate the equations of motion used to calculate reaction
dynamics on our model surface with and without dissipation, and discuss the 
specification of initial conditions on the DS.  Results are presented in Sec.\ \ref{sec:results}:
dynamics of trajectory bundles and product ratios at fixed energies with and without dissipation,
and fractional product yields as a function of dissipation parameter.
Sec.\ \ref{sec:summary} concludes. 
Details on the computations required to locate the VRI point are given in 
Appendix \ref{app:vri}.


\section{Potential energy surface}

\label{sec:pes}

In this work we investigate, using classical trajectories,
reaction dynamics on a model potential surface exhibiting 
a valley ridge inflection (VRI) point (cf.\ Figure \ref{fig:bkc_1}b).

The potential studied exhibits an index 1 saddle (`transition state', TS)
at high energy.  Trajectories are initiated on the DS  
associated with this (upper, high energy) saddle, and 
can form either of two products.  (There is also the possibility that
trajectories can exit via the high energy transition state; on sufficiently long timescales,
all trajectories [excluding a set of measure zero] will escape through this `hole' in the potential, 
provided that no energy dissipation is present.)
Downhill in energy from the upper saddle point there is another index 1 saddle point,
which forms a `ridge', which is a 
conventional transition state for the isomerization reaction
that interconverts the two products.
Between the two saddle points there is therefore 
a VRI  point \cite{Valtazanos86,Quapp98,Quapp04}.
Because the general form of the potential energy surface studied here 
is not fully symmetric with respect to the coordinate transformation $y \rightarrow -y$ (see below),
the intrinsic reaction path \cite{Fukui70} does not in fact bifurcate, so that the
location of the VRI point merely indicates the \emph{region} of the PES where, in a 
naive conventional picture,
trajectories `decide' which product well to enter  (see below). 
Most trajectories initiated on the upper DS 
do however pass through the neighborhood of the VRI point.

In addition to the trajectory studies reported in the following Section, 
we also present here results on the computation
of various theoretical constructs associated with the concept of
`reaction path' for our model surface 
\cite{Fukui70,Valtazanos86,Quapp98,Quapp04,Quapp98a,Hirsch04,Rowe82,Hoffman86}.
These are the IRC \cite{Fukui70}, 
VRI points \cite{Valtazanos86,Quapp98,Quapp04}, Newton trajectories 
\cite{Quapp98a,Hirsch04} and gradient extremal paths \cite{Rowe82,Hoffman86}.
It is worthwhile emphasizing that the various 
specific potential functions studied here do in fact
have VRI points, despite not being fully symmetric.  
Our results provide a numerical demonstration
that the mathematical conditions 
for the existence of a VRI (see below) can readily be satisfied in the
absence of symmetry.

\subsection{Model potential}

The system studied has 2 degrees of freedom (DoF), with associated coordinates
$(x, y)$.  The functional form is a modified version of a model potential previously
introduced by Carpenter (see ref.\ \onlinecite{Carpenter04}):
\begin{equation}
\label{eq:pot_1}
V(x, y) = c_0 \left( \tfrac{1}{3} x^3 - \tfrac{1}{2} \alpha x^2 \right) + 
\omega^2 y^2 \tfrac{1}{2}\left(1 - \beta x \right)
+   c_1 y^4 x + c_2 x^2 y^2 + c_3 yx^2 + c_4 x y^2 + c_5 x y + c_6 x y^3 .
\end{equation}
We fix parameter values  $\alpha = 2$,  $\beta = 2$, $c_0 = 3$, $\omega = \sqrt{3}$. 
The values of the remaining 6 parameters $c_k$, $k=1, \ldots, 6$ are then determined by 
specifying the locations and energies of the minima of the upper and lower product wells 
(6 parameters in all, obtained by solving a set of linear equations).

The upper index-1 saddle point is located at the origin $(0,0)$ with energy $V_0 =0$.
The coordinates of the lower  product well 
($y < 0$) are fixed at $(x, y) = (2.4, -1.2)$, minimum energy $v = -7.5$.
For the upper product well ($y > 0$) we take coordinates $(x, y) = (x^\ast, 1.2)$,
minimum energy $v = -6.$.   We consider 3 cases: $x^\ast = 2.00$, 
$x^\ast = 2.05$, $x^\ast = 2.10$.
Values of the coefficients $c_k$ for the 3 different cases are
given in Table \ref{table:coeff}.


\subsection{Reaction paths, bifurcations and valley ridge inflections}

Figure \ref{fig:pots_1} shows contour plots of the potentials corresponding
to values $x^\ast = 2.00$  and $x^\ast = 2.10$, respectively.
Also shown are the corresponding IRC paths connecting the upper TS with 
one of the 2 product minima.  These paths are computed in the standard way \cite{Fukui70} as
solutions of the differential equation
\begin{equation}
\deriv{\br}{s} = -\nabla V
\end{equation}
where $\br = (x, y)$ and $s$ parametrizes progress along the IRC.
Since the mass $m=1$ for our model problem, there is no distinction between 
mass-weighted and unweighted coordinates.

For $x^\ast = 2.00$ the IRC reaction path from the upper TS 
terminates at the upper minimum ($y >0$), while that for $x^\ast = 2.10$ 
terminates at the lower minimum ($y < 0$).  
(The $x^\ast = 2.05$ potential [not shown in Figure
\ref{fig:pots_1}] is also nonsymmetric; the IRC terminates at 
the upper minimum in this case.)
Table \ref{table:critical} lists coordinates and energies of 
the critical points (index-1 saddles and minima) of potential eq.\ \eqref{eq:pot_1},
computed for $x^\ast$ values $2.00$, $2.05$ and $2.10$

Additional quantities of interest are included in the contour plots of Fig.\ \ref{fig:pots_1}.
The Hessian $\calH$ is the matrix of second derivatives 
\begin{equation}
\calH = 
\begin{bmatrix} 
V_{xx}  & V_{x y} \\
V_{yx}  & V_{yy}
\end{bmatrix} 
\end{equation}
where subscripts indicate partial differentiation.
(The mass tensor/kinetic energy is by definition trivial for our model, 
as we take $m_x = m_y =1$.  When defining the Hessian it is therefore 
not necessary to consider covariant derivatives of the potential, as would be required for the
general case of a Hamiltonian having coordinate dependent
kinetic energy \cite{Tachibana81}.)
At a VRI point  \cite{Valtazanos86,Quapp98,Quapp04} (i) the Hessian matrix has a zero eigenvalue
and (ii) the gradient vector $\bu = \nabla V$ is perpendicular to the corresponding eigenvector.
As discussed in Appendix \ref{app:vri}, 
VRI points are found at the intersections of zero contours of the quantities
$\bu \cdot \text{adj}[\calH] \cdot \bu$ and $\det[\calH]$, where the adjugate
matrix $\text{adj} [ \calH ] = \det[\calH] \calH^{-1}$.

For each value $x^\ast = 2.0$ and $x^\ast = 2.1$, we show 
in Fig.\ \ref{fig:pots_1}
the zero contours of the determinant of the Hessian 
matrix $\calH$ (red) and of the quantity
$\bu \cdot \text{adj}[\calH] \cdot \bu$ (green).
Each plot exhibits a single VRI point at the
intersection of the 2 contour lines, 
close to \emph{but not actually on} the IRC path.
The locations of the VRI points for  
$x^\ast$ values $2.00$, $2.05$ and $2.10$ are listed in Table \ref{table:critical}.

In Figure \ref{fig:pots_2}a we plot a set of Newton trajectories \cite{Quapp98a,Hirsch04} 
for the case $x^\ast = 2.05$.  At every point along a Newton trajectory, the gradient
vector $\bu$ points in a fixed direction specified by a search vector $\br$ \cite{Hirsch04}.
Figure \ref{fig:pots_2}a shows Newton trajectories computed for unit search vectors
$\br = \{ \cos[\theta], \sin[\theta] \}$, for a number of angles $\theta$ sampled 
uniformly in the interval $0 \leq \theta \leq \pi$.
(In fact, the ``trajectories'' are computed as the zero contours of the function
$f = \br_{\perp} \cdot \bu$, where $\br_{\perp} = \{ -\sin[\theta], \cos[\theta] \}$ 
is a unit vector perpendicular to $\br$.)

Our results illustrate the fact that  
(complete) Newton trajectories connect 
all stationary points on the potential, 
and that bifurcations of Newton trajectories 
occur at VRI points \cite{Hirsch04}.  
These properties make Newton trajectories very useful for exploration of PES features.
Nevertheless, comparison with dynamical trajectories (see below)
shows that, at least for the potential studied here, Newton trajectories 
provide little insight into the actual reactive dynamics.

In Figure \ref{fig:pots_2}b we plot gradient extremal paths \cite{Rowe82,Hoffman86} for 
the case $x^\ast = 2.05$. 
At gradient extremal points, the gradient vector $\bu$ is an eigenvector of the Hessian
\begin{equation}
\calH \, \bu \propto \bu .
\end{equation}
The gradient extremal paths plotted are actually obtained by 
computing the zero contours of the quantity 
\cite{Hoffman86,Sun93}
\begin{equation}
\Gamma = V_{xy}(V_x^2 - V_y^2) + (V_{yy} - V_{xx}) V_x V_y .
\end{equation}

It can be seen that, in contrast to the IRC path, the gradient extremal path connects the upper and lower
index 1 saddles, even for a non-symmetric potential.  
The two index 1 saddle points are also connected by singular Newton trajectories (Fig.\ \ref{fig:pots_2}a).
However, it is also seen that gradient extremal paths can exhibit loops and turning
points, limiting their utility as models for reaction paths.


\section{Trajectory calculations: Hamiltonian, dissipation and initial conditions}
\label{sec:traj}

We study reaction dynamics using a Hamiltonian 
based on the 2 DoF potential eq.\ \eqref{eq:pot_1}.
We therefore effectively consider the dynamics on a
timescale short enough so that transfer of energy to or from other
degrees of freedom (intramolecular vibrational modes, solvent bath modes)
is negligible.  In addition, we do however (crudely) 
model the effect of additional degrees of freedom by introducing dissipation
into our model.  Explicit inclusion of additional degrees of freedom 
is left for future investigations.

\subsection{Equations of motion}

The Hamiltonian has the form:
\begin{equation}
H(x, y, p_x, p_y) = \frac{p_x^2}{2} + \frac{p_y^2}{2} + V(x, y),
\label{ham1}
\end{equation}
with potential $V(x,y)$ given by eq.\ \eqref{eq:pot_1} and Hamilton's equations of motion:
\begin{subequations}
\label{eq:hameq1}
\begin{align}
\dot{x} & = p_x,  \\ 
\dot{y} & =  p_y,  \\
\dot{p}_x & =  -\frac{\partial V}{\partial x} (x, y), \\
\dot{p}_y & =  -\frac{\partial V}{\partial y } (x, y).
\end{align}
\end{subequations}
The effects of dissipation are modelled by adding a simple damping term to equations 
of motion \eqref{eq:hameq1} as follows:
\begin{subequations}
\label{eq:hameq_diss1}
\begin{align}
\dot{x} & =  p_x, \\ 
\dot{y} & =  p_y, \\
\dot{p}_x & =  -\frac{\partial V}{\partial x} (x, y)- \gamma_x p_x, \\
\dot{p}_y & =  -\frac{\partial V}{\partial y } (x, y) - \gamma_y p_y.
\end{align}
\end{subequations}
for some $\gamma_x, \, \gamma_y >0$, so that the
kinetic energy monotonically decreases along the trajectory.
We set  $\gamma_x= \gamma_y \equiv \gamma$
and study the effects of dissipation for a range of $\gamma$ values $0 \leq \gamma \leq 1$.
In the present calculations, random thermal fluctuations (e.g., Langevin dynamics \cite{Zwanzig01}) are not
considered.

\subsection{Initial conditions}

Trajectories are initiated on the phase space DS associated with the 
transition state located at the high energy saddle point. 
A normal form \cite{Waalkens08} of degree 10 is computed, and 
the dividing surface sampled using a grid in phase
space at a specified energy \cite{Collins11}. 
We integrate trajectories and compute  product fractional yield 
(equivalently, branching ratio) as 
a function of time.

In order to decide  whether a trajectory is in the upper or lower product well,
we define a plane in phase space tangent to the dividing surface 
separating the two products; this surface is 
computed from a normal form constructed at the lower (ridge) saddle point.
The sign of the standard inner product of the displacement vector of
a phase point from the lower saddle with a vector normal to the tangent plane then
determines the well to which the point is assigned.
Trajectories are stopped if they re-cross
the upper saddle DS but few such cases were observed (for
the integration times used), and none for any nonzero
values of the dissipation factor $\gamma$ (see below).

The effects of dissipation are modelled by integrating eq.\ \eqref{eq:hameq_diss1}
for a range of values of the dissipation parameter $\gamma$.
Initial conditions for trajectory  calculations incorporating dissipation 
are  sampled on the DS
in the usual way using Hamiltonian \eqref{eq:hameq1} for a 
fixed value of the initial energy.


\section{Results}
\label{sec:results}

\subsection{Reaction dynamics without dissipation}

Figure \ref{fig:traj_e0_1} shows the behavior of bundles of 64 trajectories initiated
on  the DS at the upper TS, with energy $E = 0.1$ above the
saddle energy.  (Recall that the energy of the upper saddle is $E=0$, 
the energy of the lower saddle is $E \sim -4.0$, while
the product minima are at energies $E=-6.0$ and $E=-7.5$, repectively.)
We show results for the 2 cases $x^\ast = 2.0$ and $x^\ast = 2.1$, respectively.
Trajectories are integrated for $t_{\text{max}} = 4$ time units, a time comparable to the natural 
period for motion in either well.

Figure \ref{fig:traj_e0_01} shows corresponding trajectory bundles at $E=0.01$ 
above the saddle energy.  These lower energy trajectories are integrated for
longer times, up to $t_{\text{max}} = 8$.

All trajectories initially
collide with the hard wall of the potential that is directly `downhill' from the upper TS.  
This collision is
followed by a number of more or less `coherent' oscillations 
of the trajectory bundle between product wells,
with concurrent dephasing.
The well occupancies (product yields) shown in Figs \ref{fig:traj_e0_1} and \ref{fig:traj_e0_01}
clearly demonstrate the coherent short-time behavior of the trajectory bundles, 
and the dramatic effect on well occupancies brought about by apparently
minor changes in potential topography.

Changing the location of the upper minimum, specifically the $x$-coordinate $x^\ast$,
also changes the curvature of the potential in the vicinity 
of the `hard wall' encountered by trajectories after they have rolled 
downhill from the upper TS.  This change in curvature in turn affects the 
direction in which the trajectory bundle is predominantly `reflected' by the hard wall, as can be seen
from the time-dependent product yields (fractions)
shown in Figures \ref{fig:traj_e0_1} and \ref{fig:traj_e0_01}

Figure \ref{fig:WellEn0.01t100} shows well occupancies for the 2 cases $x^\ast = 2.0$ and $x^\ast = 2.1$
for  $0 \leq t \leq t_{\text{max}} =100$.  Even at $t = t_{\text{max}}$, the ratio of well occupancies
apparently has not converged to an asymptotic (steady state) value. That is, nontrivial 
isomerization dynamics is still occuring.
These fluctuations may however reflect the finite size of the trajectory ensemble used
in our calculations.  Figure \ref{fig:WellEn0.01t100} also shows that 
the cumulative fraction of trajectories that escape (recross the upper DS) is 
small but non-negligible for both cases.

We next consider the addition of dissipative damping, which 
ensures that the branching ratio becomes well defined at relatively short times.

\subsection{Reaction dynamics with dissipation}

Figure \ref{fig:traj_diss_1} shows 2 trajectory bundles at $E = 0.01$
with the relatively large  dissipation factor $\gamma = 0.5$
(64 trajectories per bundle).
The trajectories drop into one of the wells within approximately 4 time units, and,
as anticipated for the highly dissipative case, 
the predominant product obtained is determined
by the IRC path from the upper TS.
Note that the product ratio inverts between the two cases, which differ \emph{only}
in the value of $x^\ast$.

Figure \ref{fig:traj_diss_2} shows the behavior of trajectory bundles 
at $E=0.01$ for a smaller dissipation factor $\gamma = 0.25$ 
together with corresponding well occupancies, for $0 \leq t \leq t_{\text{max}} = 5$.
For $x^\ast = 2.1$, the
product ratio at long times is now reversed with respect to 
the value for  $\gamma = 0.5$; for the lower dissipation parameter,
trajectories are able to cross the ridge separating products
one more time (on average) before losing energy and becoming trapped in one or the other well.

These results suggest the interesting possibility that the branching ratios
might exhibit a \emph{non-monotonic} dependence on dissipation parameter.
This question is explored below.

\subsection{Product ratios as a function of dissipation parameter}

We now examine systematically the behavior of 
fractional product yields as a function of the dissipation parameter.
The branching ratio is given in terms of the fraction of 
trajectories which are in either of the two wells after the system has settled down  
and trajectories no longer have sufficient energy to cross the ridge.
As the dissipation factor $\gamma$ becomes smaller, it is necessary to follow trajectory
ensembles for longer and longer times to determine asymptotic product ratios.

Note that, although our trajectory calculations examine the fate of ensembles of
trajectories initiated on the DS at a fixed time, the branching ratios we compute
are nevertheless equally applicable to the situation in  which a steady stream of reactants
passes over the upper TS.

We compute fractional product yields for a range of 
dissipation parameter $0.01 \leq \gamma \leq 1$.
We have checked that our results are converged both with respect to
the trajectory run time $t_{\text{max}}$ and the size of the ensemble.

Figure \ref{fig:diss} shows the fraction of trajectories
in the lower well as a function of dissipation parameter
$\gamma$ for $x^\ast = 2.0$ (Fig.\ \ref{fig:diss}a), $x^\ast = 2.05$ (Fig.\ \ref{fig:diss}b) and 
$x^\ast = 2.1$ (Fig.\ \ref{fig:diss}c) for $E = 0.01$.  For each
case, the fraction of given product is a highly structured non-monotonic
function of the dissipation parameter $\gamma$.  
It is moreover striking that minor variations in the value of $x^\ast$  lead to 
noticeably different dependence of yield on $\gamma$.
Figure \ref{fig:diss}d shows the fractional yield for $x^\ast = 2.05$ at the higher energy
$E = 0.1$ above threshold; the behavior 
is very similar to that seen at the lower energy $E=0.01$.

The nontrivial dependence of branching ratios on $\gamma$ has its origin in the interplay
between the almost coherent ridge crossing dynamics of the trajectory 
bundle and the dissipative loss of kinetic energy, leading to trapping of trajectories
in one or the other well.
The dissipation rate sets the timescale on which trajectories 
settle into their final associated product wells.

This  interpretation of the branching ratio results is
confirmed by examining the dynamics of trajectory bundles in more detail.
For example, Figure \ref{fig:segments} shows time-dependent 
fractional yields and trajectory segments ($t_1 \leq t \leq t_2$) for
the intermediate case $x^\ast = 2.05$ at initial energy $E = 0.01$ for 2 dissipation parameters,
$\gamma = 0.13$ and $\gamma = 0.1$, respectively.
Ensemble trajectory segments for the ensemble are shown with $t_1 = 8.5$, $t_2 =9.0$.
With these values of $t_1$ and $t_2$ it can be seen that 
for the larger dissipation parameter 
the final product well is already determined for all trajectories in the ensemble,
while reducing the dissipation parameter slightly
allows some additional ridge crossing, changing the branching
ratio significantly.

The coherent crossing of the central ridge causes trajectories to 
sample one product well and then the other in an oscillatory fashion.  
Similar phenomena have been seen in trajectory simulations of a number 
of unimolecular reactions of polyatomic systems \cite{Goldman11,Nummela02}, 
suggesting that the present behavior might persist on at 
least some higher dimensional potential energy surfaces.  
This question is under active investigation.

\newpage

\section{Summary and conclusions}
\label{sec:summary}

We have studied reaction dynamics on a model potential energy surface 
exhibiting post-transition state bifurcation in the vicinity of
a valley ridge inflection point.  
Bundles of trajectories initiated on the dividing surface associated with a high energy TS 
are followed into the region of the PES where bifurcation of the reaction path occurs, 
and the subsequent dynamics studied. 
We have computed fractional yields for 
products reached after the VRI region is traversed, both with and
without dissipation.
It is found that 
apparently minor variations in the potential lead to significant changes in the reaction
dynamics.  Moreover, the branching ratio depends in a complicated and highly non-monotonic fashion
on the dissipation parameter.

For the model considered here,  
the dynamics proceeds on at least two timescales.  First, on short times, the
bundle of trajectories `reflects' off a hard wall that is opposite the
high energy TS through which it enters the reaction zone.  
After this collision, a highly non-statistical and time-dependent 
population ratio of products is established,
whose value depends on the direction in which trajectories are reflected by
the (asymmetric) potential wall. These initial
nonstatistical populations then relax on a somewhat longer timescale to 
yield the observed product ratio.  
During this phase of the reaction, there is 
the possibility of competition between IVR and other mechanisms 
for removing vibrational energy from active degees of freedom.
Introducing dissipation into the model sets the timescale on which the branching
ratio is determined.

One would expect that for real chemical systems, reactions 
in condensed phases would be characterized by 
higher dissipation rates \cite{Crim12}  
and that some control of collision-induced dissipation might be attainable 
through the use of supercritical fluids at variable pressure \cite{Schwarzer96}.  
However, to our knowledge, such techniques have not yet been applied to any reaction for 
which the existence of a chemically significant VRI has been established.

Overall, we find that dynamics in the vicinity of the VRI point on the potential
play essentially no role in
determining the product ratio, except in the highly dissipative regime.
Extension of these investigations to more
realistic theoretical models of reactions involving VRI points (see, for example,
refs \onlinecite{Arnold00,Mann02,Mann02a,Quapp11,Quapp12})
is a topic of current research.

\acknowledgments

We are grateful to Dr.\ Zeb Kramer for his 
penetrating comments on the manuscript.

PC and SW acknowledge the support of the Office of Naval Research 
(Grant No.~N00014-01-1-0769). 
PC, BKC and SW acknowledge the support 
of the UK Engineering and Physical Sciences Research Council (Grant No.\ EP/K000489/1).
The work of GSE is supported by the US National Science Foundation under 
Grant No.\ CHE-1223754.


\appendix

\section{Location of VRI points}
\label{app:vri}

The Hessian matrix $\calH$ is a real symmetric matrix, and has 
2 real eigenvalues and associated orthonormal eigenvectors.
Let the eigenvalues and eigenvectors of $\calH$ be denoted $\lambda_\alpha$ and $\bv_\alpha$,
respectively, $\alpha = 1,2$.  The potential gradient vector $\bu = \nabla V$ is then
decomposed as follows:
\begin{equation}
\bu = \sum_{\alpha} \bv_\alpha c_\alpha .
\end{equation}
Defining the \emph{adjugate} matrix $\text{adj}[\calH] \equiv \det[\calH] \,  \calH^{-1}$,
we have
\begin{equation}
\bu \cdot \text{adj}[\calH] \cdot \bu = \sum_{\alpha} \lambda_{\alpha^\prime} \, c_\alpha^2
\end{equation}
where $\alpha^\prime = 2$ if $\alpha =1$, and vice versa.
The quantity $\bu \cdot \text{adj}[\calH] \cdot \bu$ is the 
second-order variation in the potential along a vector perpendicular to $\bu$ 
(having the same length).
The condition
\begin{equation}
\label{eq:vri_1}
\bu \cdot \text{adj}[\calH] \cdot \bu = 0
\end{equation}
therefore implies
\begin{equation}
\label{eq:vri_2}
c_1^2 \lambda_2 = - c_2^2 \lambda_1.
\end{equation}

At the VRI point, there is an eigenvector perpendicular to the
gradient vector $\bu$ with associated eigenvalue zero. 
Therefore, at the VRI point, both condition \eqref{eq:vri_1}
and the condition
\begin{equation}
\det[\calH] = \lambda_1 \lambda_2 =0
\end{equation} must hold.  We therefore find the VRI point(s)
numerically by locating the intersection(s) of the zero contours of $\det[\calH]$
and $\bu \cdot \text{adj}[\calH] \cdot \bu$.


\def\cprime{$'$}

\newpage

\begin{table}[H]
\begin{center}
\begin{tabular}{|c|c|c|c|c|c|c|}
\hline\hline
$x^\ast$ & $c_1$  & $c_2$  & $c_3$  & $c_4$  & $c_5$  & $c_6$ \\
\hline
 2.05 & 0.622088 & 0.249578 & 0.401924 & 0.0040275 & -0.459044 & -0.106126 \\
 2.0 & 0.619384 & 0.318432 & 0.484549 & -0.131496 & -0.63717 & -0.0986087 \\
 2.1 & 0.625518 & 0.183133 & 0.32219 & 0.136952 & -0.286571 & -0.111012 \\
 \hline
 \end{tabular}
 \end{center}
 \caption{\label{table:coeff}  Coefficients $c_k$ for the potential eq.\ \eqref{eq:pot_1}; 
 sets of coefficients listed correspond 
 to different values of the coordinate $x^\ast$ specifying the location of the 
 upper minimum in the potential.}
 \end{table}

 \begin{table}[H]
 \begin{center}
 \begin{tabular}{|c|c|c|c|c|}
 \hline 
$x^\ast$ & Critical point & $x$ & $y$ & $E$ \\
\hline
2.0 (U) & Upper saddle & 0 & 0 & 0 \\
 & Lower saddle (ridge) & 1.983& 0.093 & -3.969 \\
 & Upper minimum & 2 & 1.20 & - 6.00 \\
 & Lower minimum & 2.4 & -1.20 & -7.50 \\
 & VRI point & 0.508 & 0.020 & -0.647 \\
 \hline
 2.05 (U) & Upper saddle & 0 & 0 & 0 \\
  & Lower saddle (ridge) & 1.985 & 0.097 & -3.967 \\
  & Upper minimum & 2.05 & 1.20 & -6.00 \\
  & Lower minimum & 2.4 & -1.20 & -7.50 \\
  & VRI point & 0.524 & 0.022 &  -0.683 \\
 \hline
2.1 (L) & Upper  saddle & 0 & 0 & 0 \\
 & Lower saddle (ridge) & 1.987 & 0.101 & -3.964 \\
 & Upper minimum & 2.1 & 1.20 & -6.00 \\
 & Lower minimum & 2.4 & -1.20 & -7.50 \\
 & VRI point & 0.542 & 0.025 & -0.723 \\
 \hline
 \end{tabular}
 \end{center}
 \caption{\label{table:critical} Coordinates $(x, y)$ and
 energies $E$ of critical points of potential eq.\ \eqref{eq:pot_1},
 computed for different values of coordinate $x^\ast$.  Coordinates 
 and energies of VRI points  are also listed.  We indicate whether the 
 IRC initiated at the upper TS terminates at the upper (U) or lower (L) minimum.} 
 \end{table}

\newpage

\section*{Figure Captions}

                \begin{figure}[H]
                          \caption{\label{fig:bkc_1} 
                          (a) Typical reaction profile of two consecutive transition states (T1, T2) 
                          linking starting material (S) with the intermediate (I) 
                          and product (P). (b) PES featuring a VRI as an alternative mechanistic 
                          situation featuring two transition states, but no intermediate.
                          (Fig.\ 1 of ref.\ \onlinecite{Rehbein11}.)}
                 \end{figure}
                 

\begin{figure}[H]
\caption{Contour plots of the potentials corresponding
to values $x^\ast = 2.00$  and $x^\ast = 2.10$, respectively.
Also shown are the corresponding IRC paths (blue) connecting the upper TS with 
one of the 2 product minima.  The locations of the 
saddle points and minima are given in Table \ref{table:critical}.
For each value $x^\ast = 2.0$ and $x^\ast = 2.1$, we plot zero contours of the determinant of the Hessian 
matrix $\calH$ (red) and of the quantity
$\bu \cdot \text{adj}[\calH] \cdot \bu$ (green).
Each plot exhibits a single VRI point at the
intersection of the 2 contour lines, 
close to but not actually on the IRC path.  Locations 
and energies of the VRI points are given in Table 
\ref{table:critical}.
}
\label{fig:pots_1}
\end{figure}


\begin{figure}[H]
\caption{(a)  Newton trajectories (magenta/green) for 
the case $x^\ast = 2.05$. The green trajectories are singular trajectories 
connecting critical points to the VRI point on the potential.
(b) Gradient extremal paths (magenta) for 
the case $x^\ast = 2.05$. 
At gradient extremal points, the gradient vector $\bu$ is an eigenvector 
of the Hessian $\calH$.  The IRC (blue) is also included for comparison.
}
\label{fig:pots_2}
\end{figure}


\begin{figure}[H]
\caption{Trajectory bundles and fractional product yields for $x^\ast = 2.0$ and $x^\ast = 2.1 $ at energy $E=0.1$.
No dissipation ($\gamma = 0$).}
\label{fig:traj_e0_1}
\end{figure}


\begin{figure}[H]
\caption{Trajectory bundles and fractional product yields for $x^\ast = 2.0$ and $x^\ast = 2.1 $ at energy $E=0.01$.
No dissipation ($\gamma = 0$).}
\label{fig:traj_e0_01}
\end{figure}


\begin{figure}[H]
\caption{Fractional product yields for (a) $x^\ast = 2.0$ and (b) $x^\ast = 2.1 $ at energy $E=0.01$ for 
  times $0 \leq t \leq t_{\text{max}} =100$.  The dashed blue line
  shows the cumulative fraction of trajectories recrossing the upper DS.}
\label{fig:WellEn0.01t100}
\end{figure}


\begin{figure}[H]
\caption{Trajectory bundles and fractional product yields for $x^\ast = 2.0$ and $x^\ast = 2.1 $.
Initial energy $E=0.01$, dissipation parameter  $\gamma = 0.5$.}
\label{fig:traj_diss_1}
\end{figure}


\begin{figure}[H]
\caption{Trajectory bundles and fractional product yields for $x^\ast = 2.0$ and $x^\ast = 2.1 $.
Initial energy $E=0.01$, dissipation parameter  $\gamma = 0.25$.}
\label{fig:traj_diss_2}
\end{figure}


\begin{figure}[H]
\caption{Fraction of trajectories in the lower well 
versus dissipation parameter $\gamma$.  (a) $x^\ast = 2.0$, $E = 0.01$; 
(b) $x^\ast = 2.05$, $E = 0.01$; (c) $x^\ast = 2.1$, $E = 0.01$; 
(d) $x^\ast = 2.05$, $E = 0.1$.
\label{fig:diss}}
\end{figure}


\begin{figure}[H]
\caption{Time-dependent fractional product yields and trajectory segments,
$t_1 \leq t \leq t_2$, $t_1 =8.5$, $t_2 =9.0$, $x^\ast= 2.05$, initial energy 
$E = 0.01$.
(a) Trajectory segments, $\gamma = 0.13$; 
(b) Fractional yield vs $t$, $\gamma = 0.13$; 
(c) Trajectory segments, $\gamma = 0.1$; 
(d) Fractional yield vs $t$, $\gamma = 0.1$.
\label{fig:segments}}
\end{figure}

\newpage


                \begin{figure}[H]
                      \centering
                      \includegraphics[angle=0,width=17.0cm]{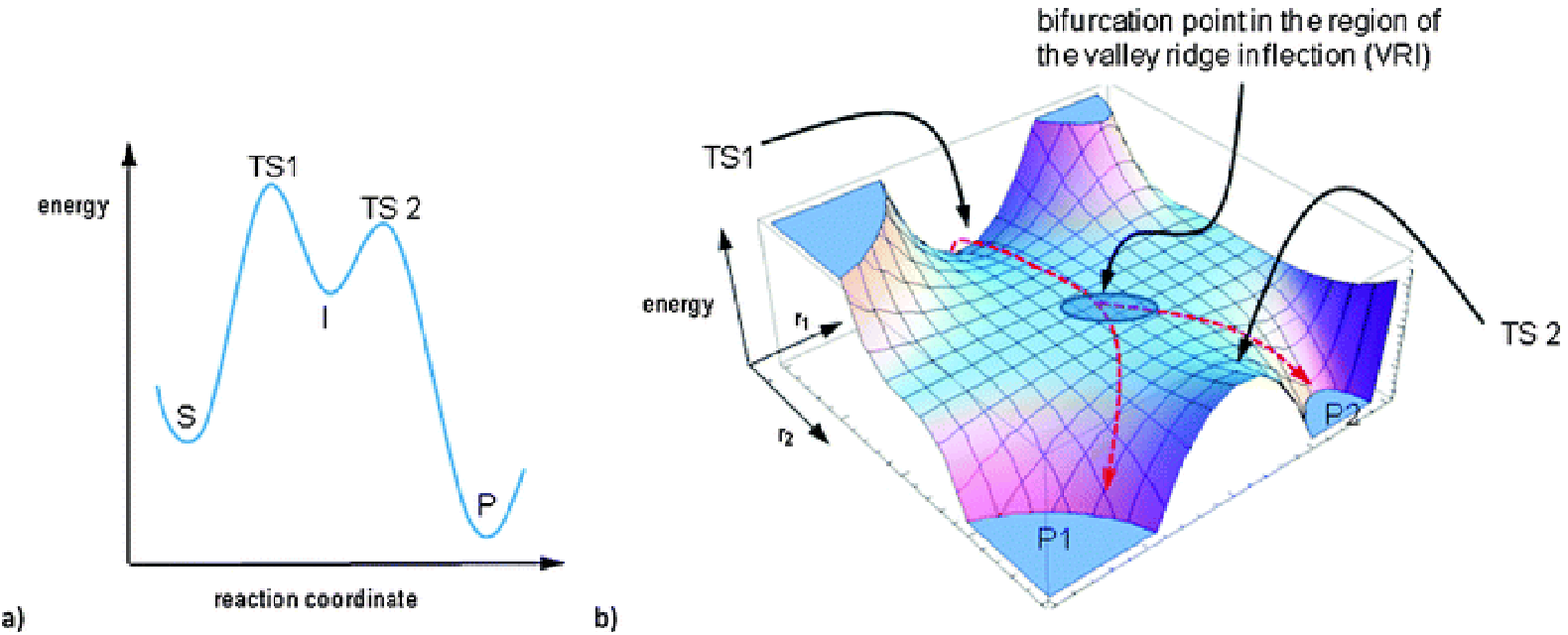}
                 \end{figure}
                 
 \vspace{0.5in}
 FIGURE 1
                 

\newpage

\begin{figure}[H]
\begin{center}
\includegraphics[angle=0,width=8.0cm]{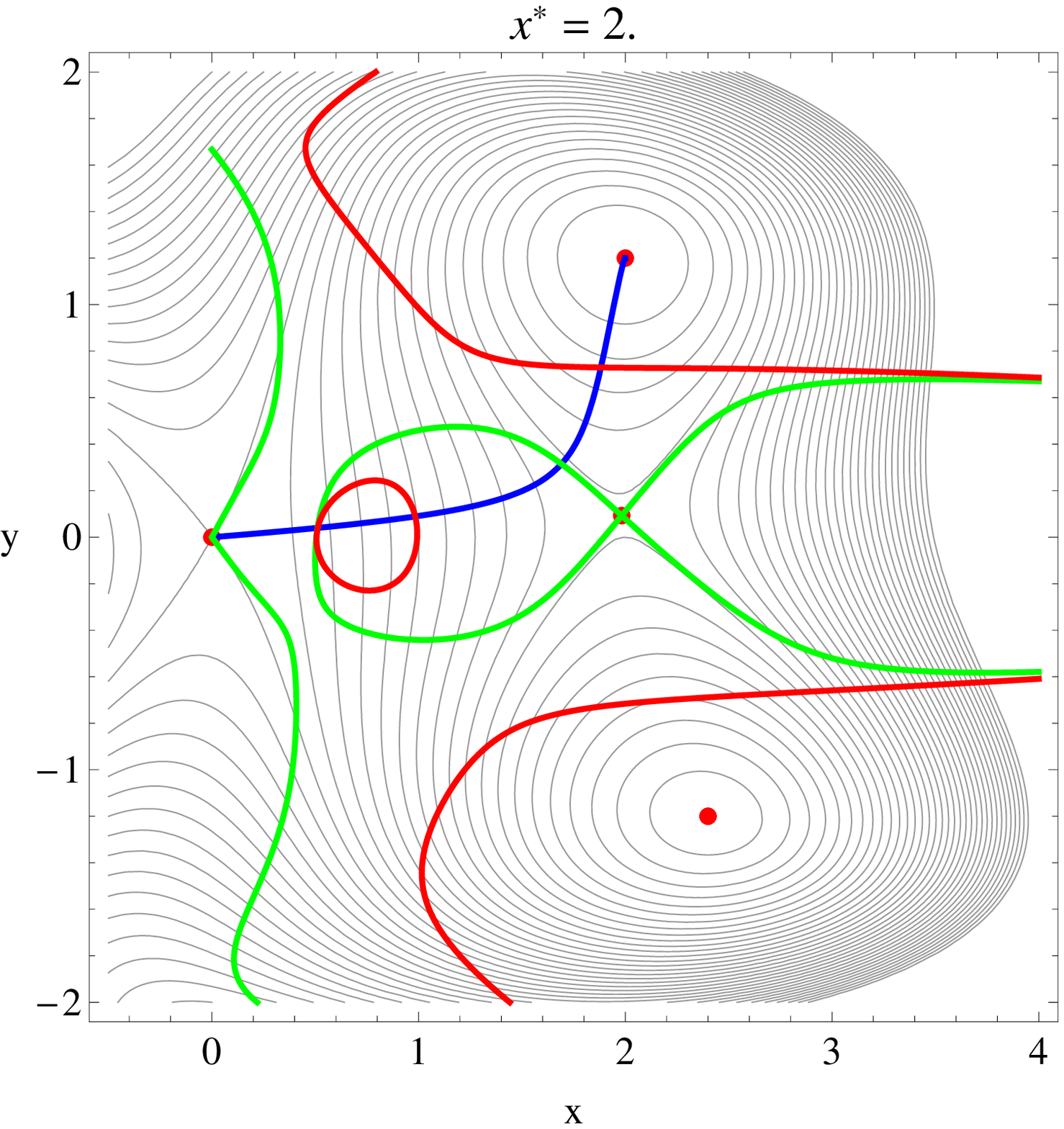} 
\includegraphics[angle=0,width=8.0cm]{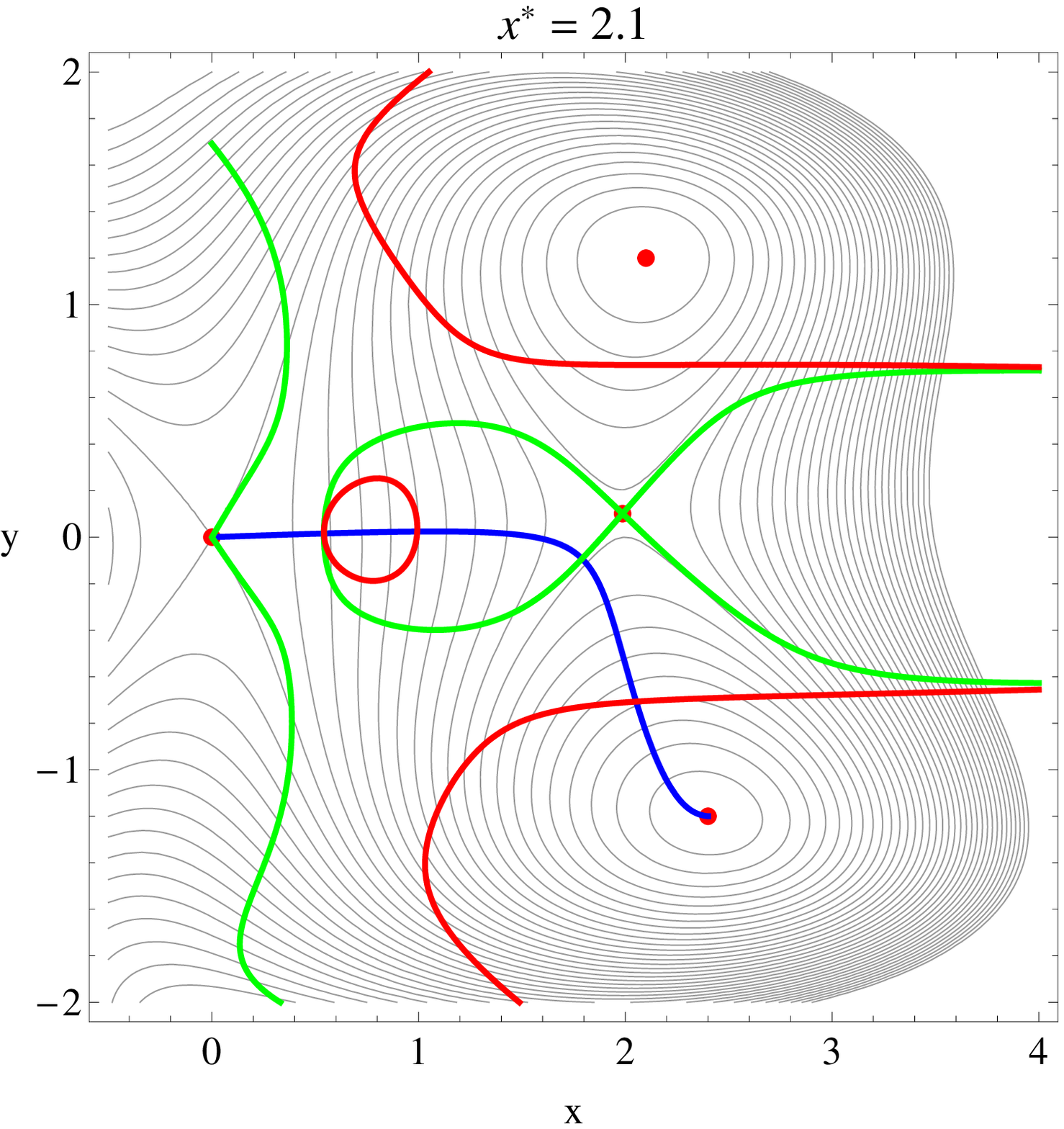}
\end{center}
\end{figure}

 \vspace{0.5in}
 FIGURE 2


\newpage

\begin{figure}[H]
\begin{center}
\includegraphics[angle=0,width=8.0cm]{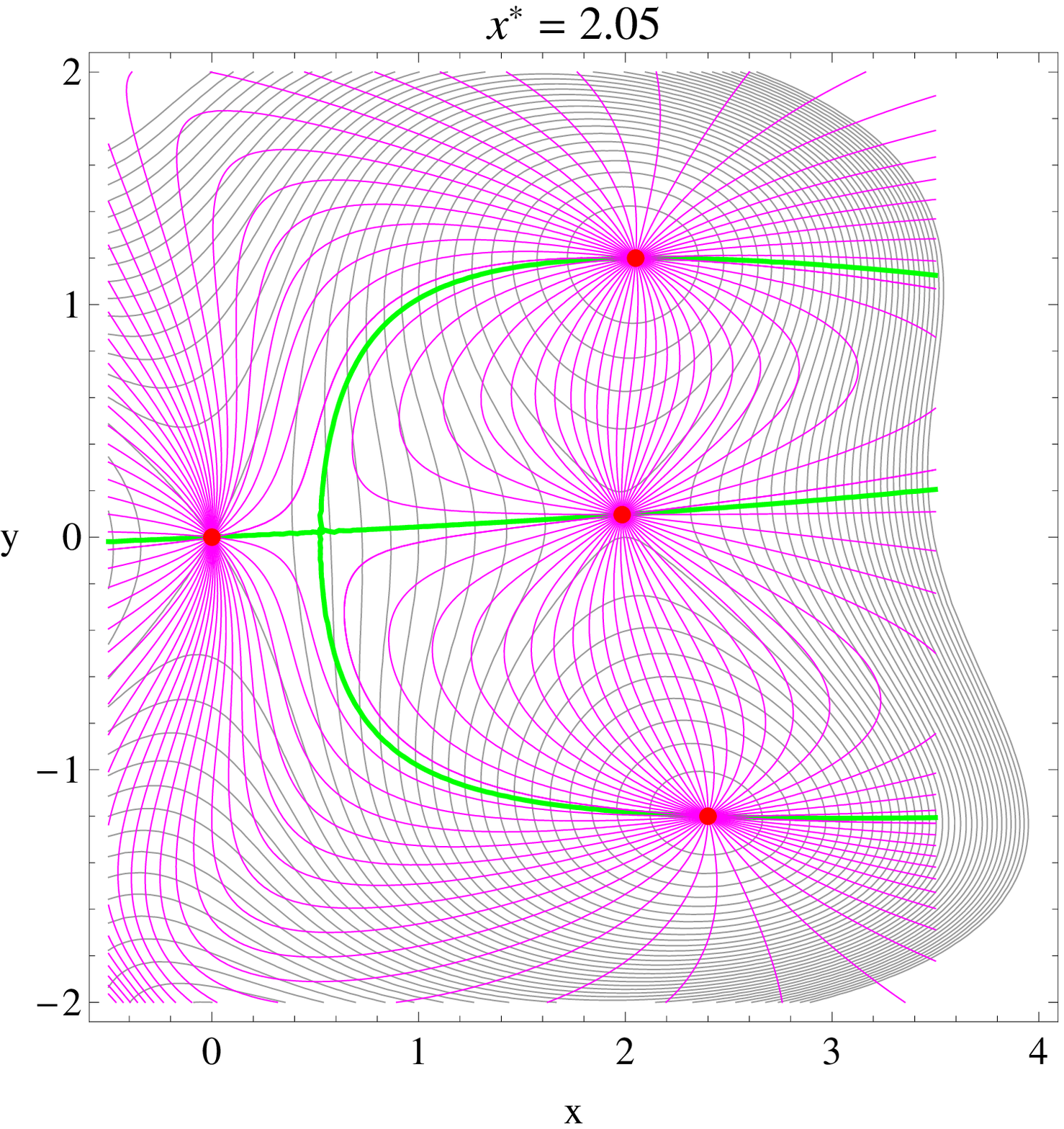} 
\includegraphics[angle=0,width=8.0cm]{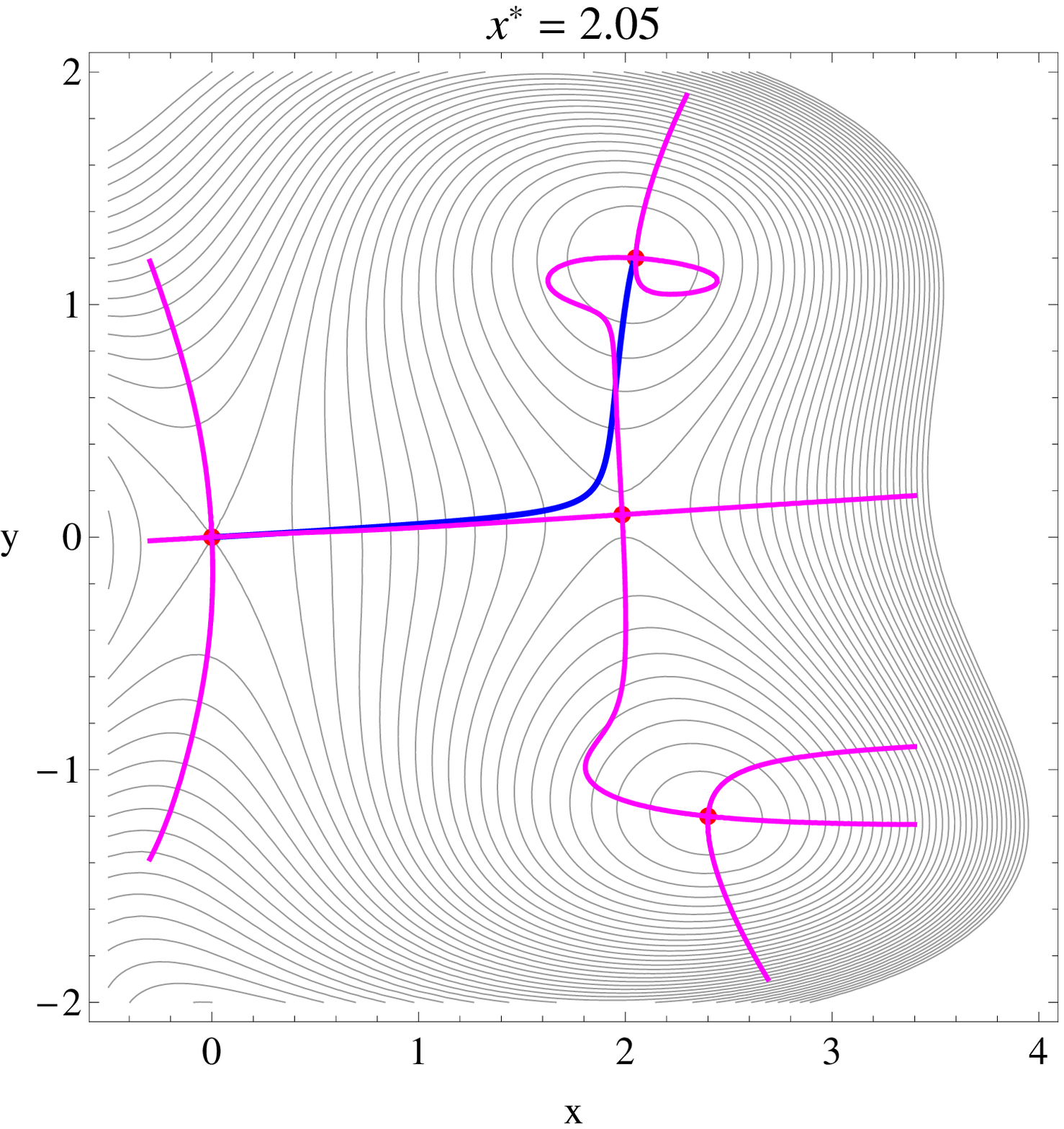} 
\end{center}
\end{figure}

 \vspace{0.5in}
 FIGURE 3


\newpage

\begin{figure}[H]
\begin{center}
\includegraphics[angle=0,width=8.0cm]{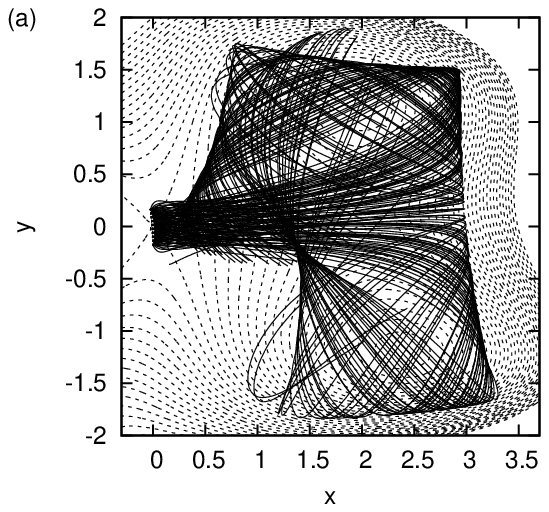} 
\includegraphics[angle=0,width=8.0cm]{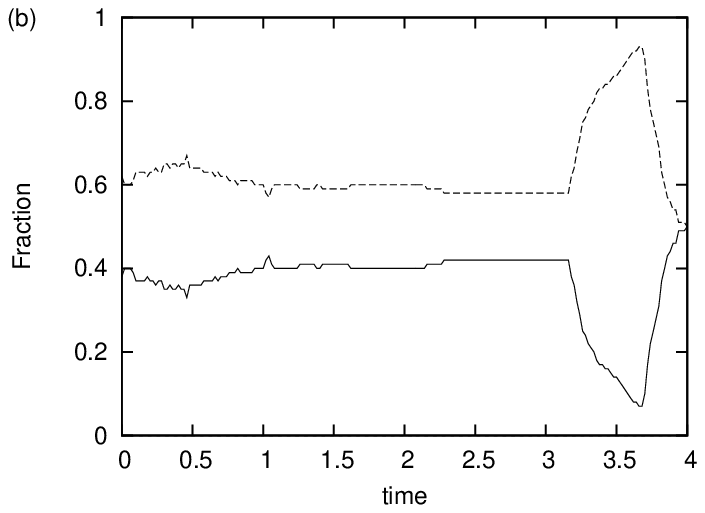} \\
\includegraphics[angle=0,width=8.0cm]{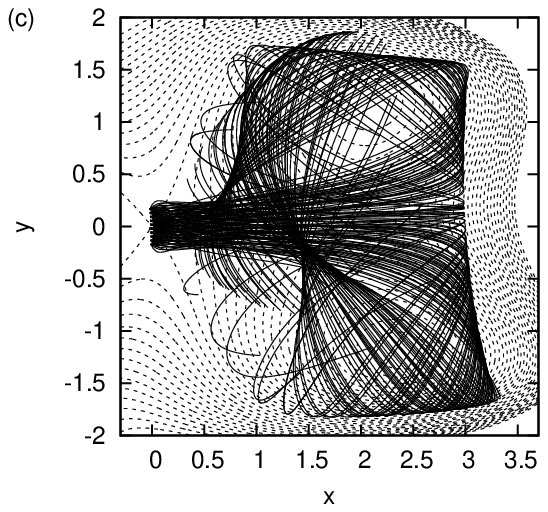} 
\includegraphics[angle=0,width=8.0cm]{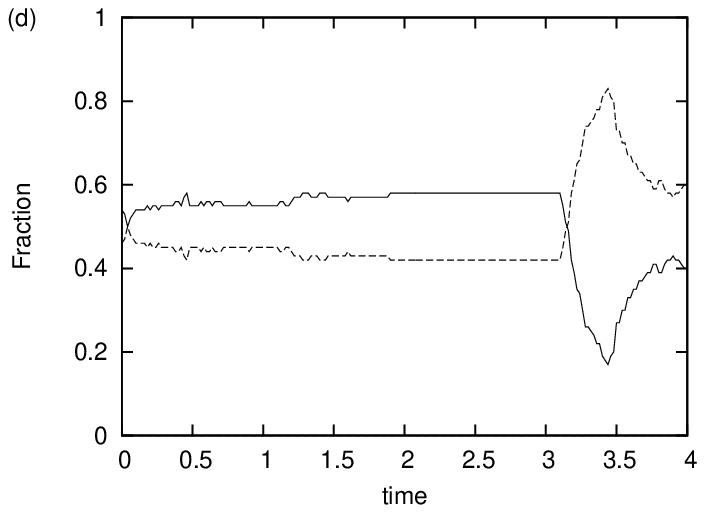} 
\end{center}
\end{figure}

 \vspace{0.5in}
 FIGURE 4


\newpage

\begin{figure}[H]
\begin{center}
\includegraphics[angle=0,width=8.0cm]{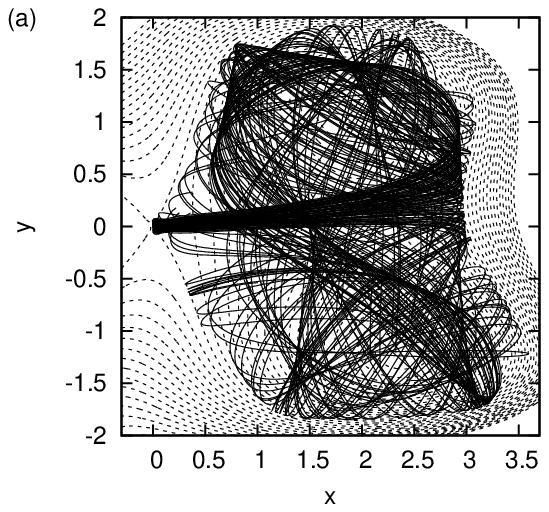} 
\includegraphics[angle=0,width=8.0cm]{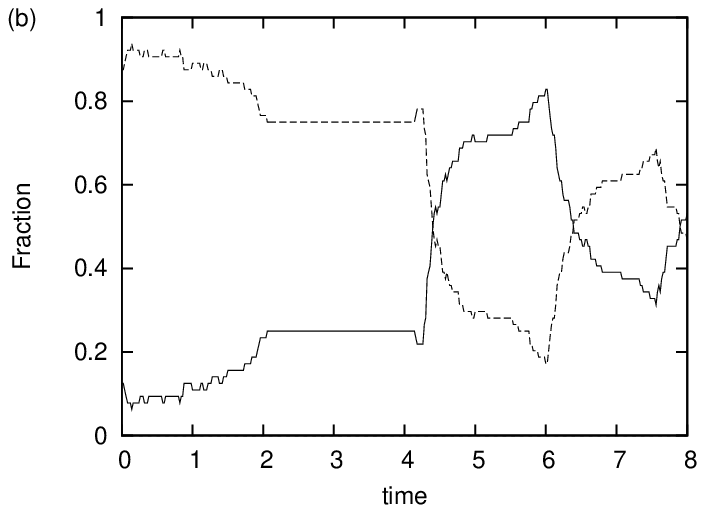} \\
\includegraphics[angle=0,width=8.0cm]{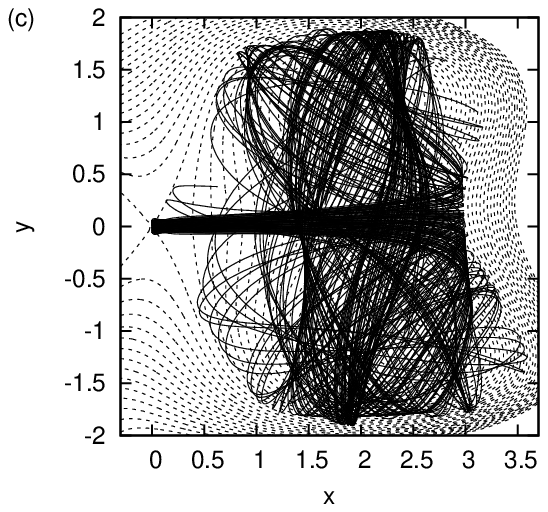} 
\includegraphics[angle=0,width=8.0cm]{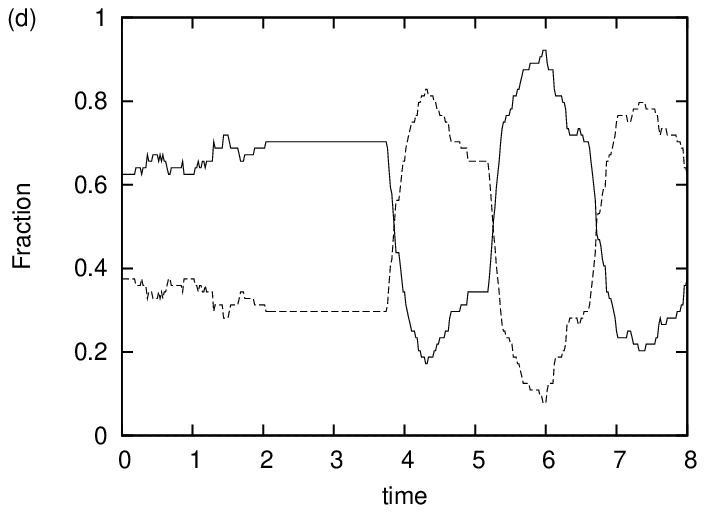} 
\end{center}
\end{figure}

\vspace{0.5in}
 FIGURE 5


\newpage

\begin{figure}[htbp!]
\begin{center}
\includegraphics[angle=0,width=8.0cm]{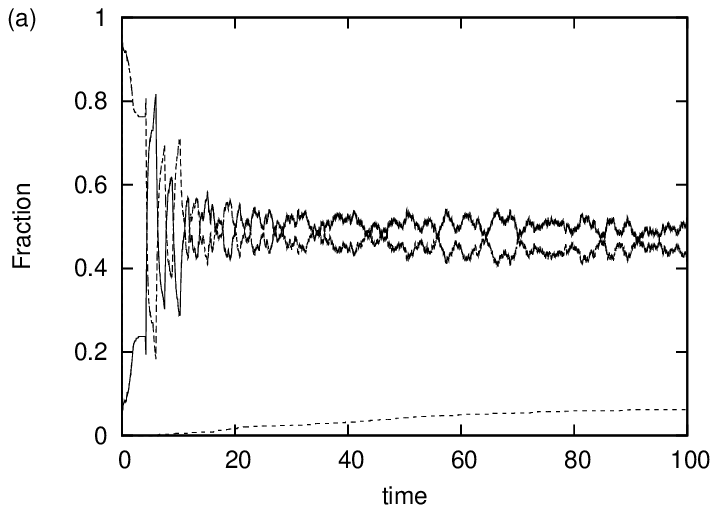} 
\includegraphics[angle=0,width=8.0cm]{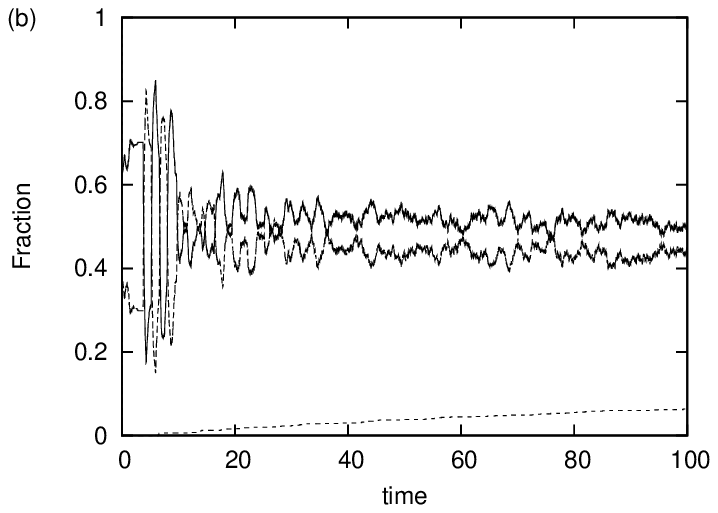} \\
\end{center}
\end{figure}

\vspace{0.5in}
 FIGURE 6


\newpage

\begin{figure}[H]
\begin{center}
\includegraphics[angle=0,width=8.0cm]{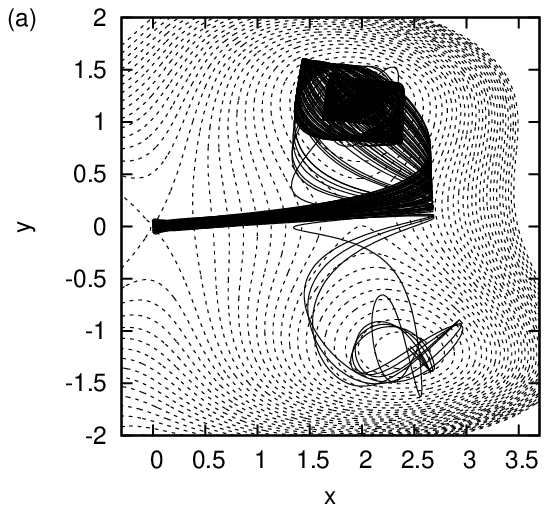} 
\includegraphics[angle=0,width=8.0cm]{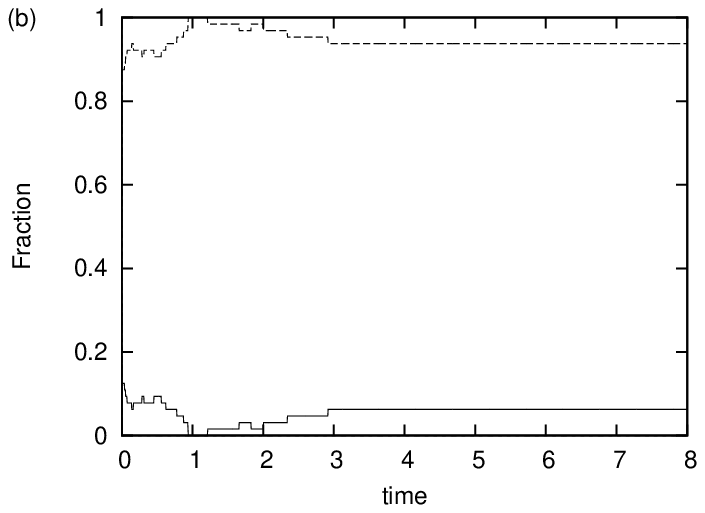} \\
\includegraphics[angle=0,width=8.0cm]{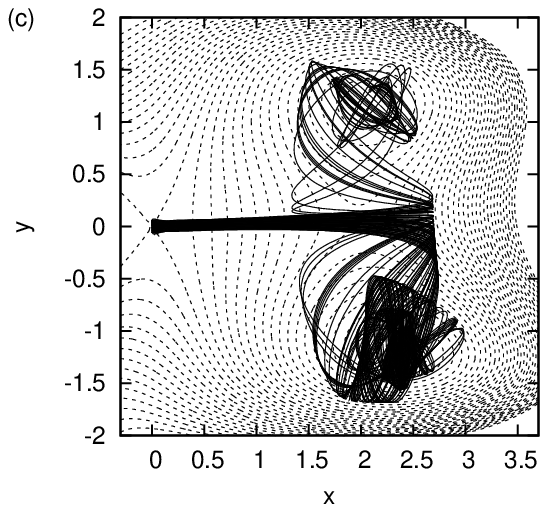} 
\includegraphics[angle=0,width=8.0cm]{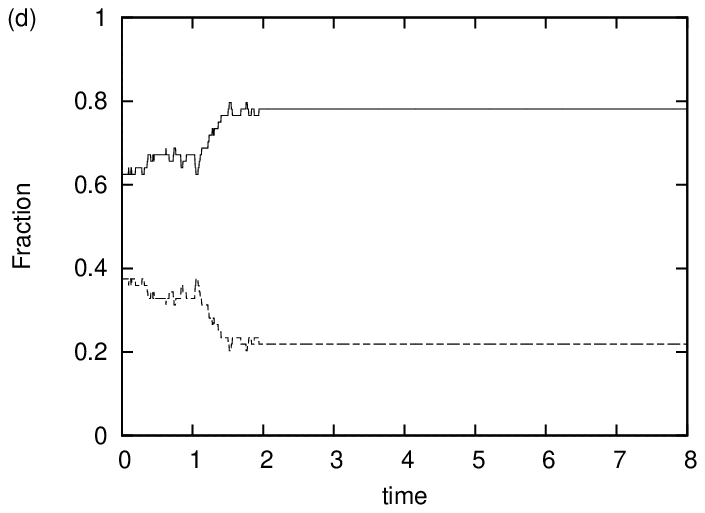} 
\end{center}
\end{figure}

\vspace{0.5in}
 FIGURE 7


\newpage

\begin{figure}[H]
\begin{center}
\includegraphics[angle=0,width=8.0cm]{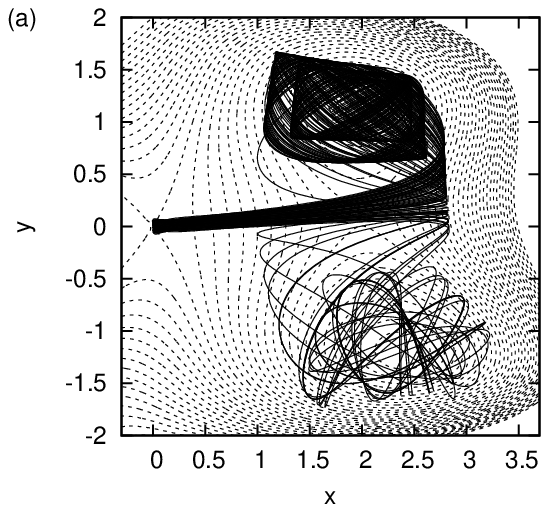} 
\includegraphics[angle=0,width=8.0cm]{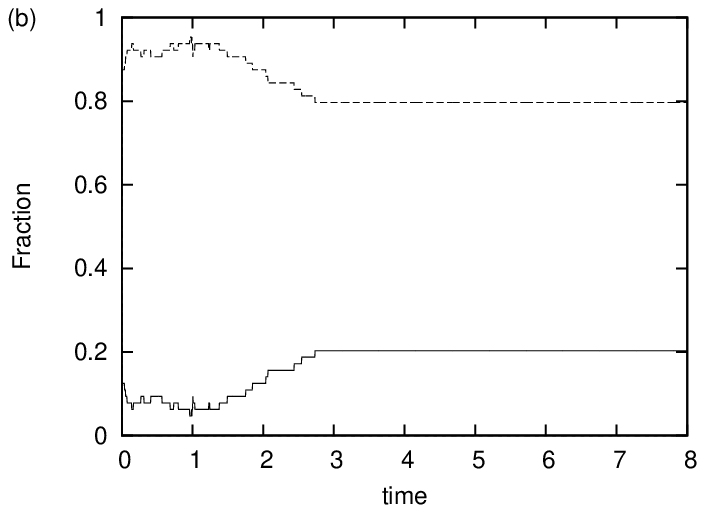} \\
\includegraphics[angle=0,width=8.0cm]{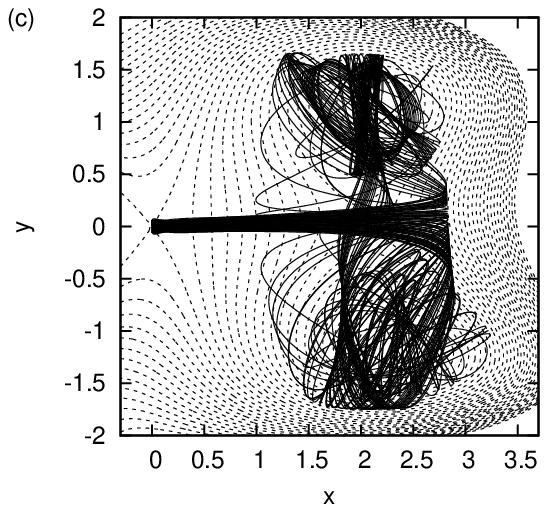} 
\includegraphics[angle=0,width=8.0cm]{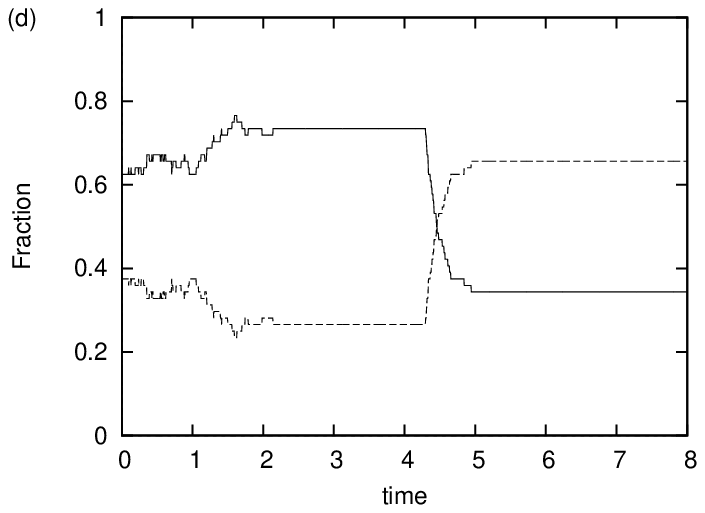} 
\end{center}
\end{figure}

\vspace{0.5in}
 FIGURE 8


\newpage 

\begin{figure}[H]
\begin{center}
\includegraphics[angle=0,width=8.0cm]{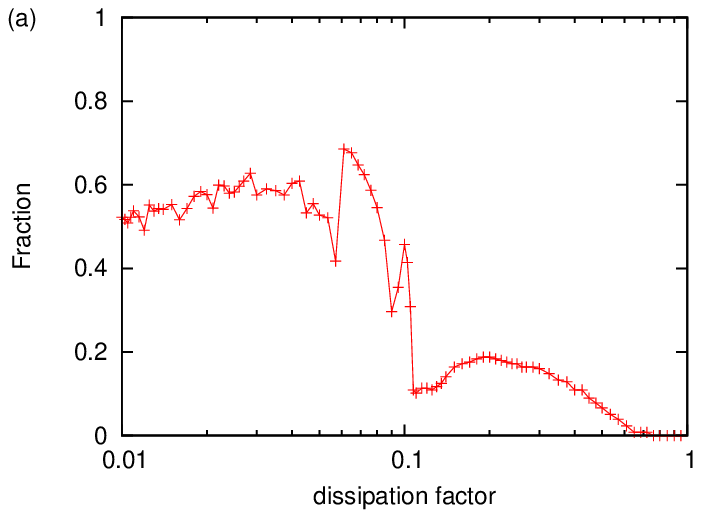} 
\includegraphics[angle=0,width=8.0cm]{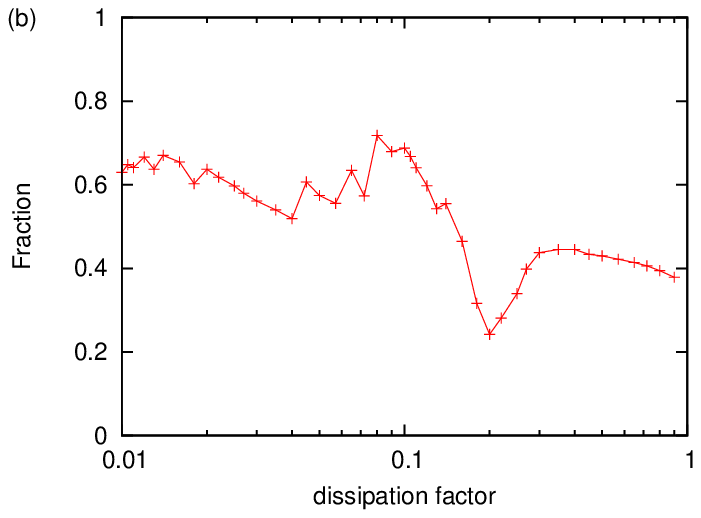} \\
\includegraphics[angle=0,width=8.0cm]{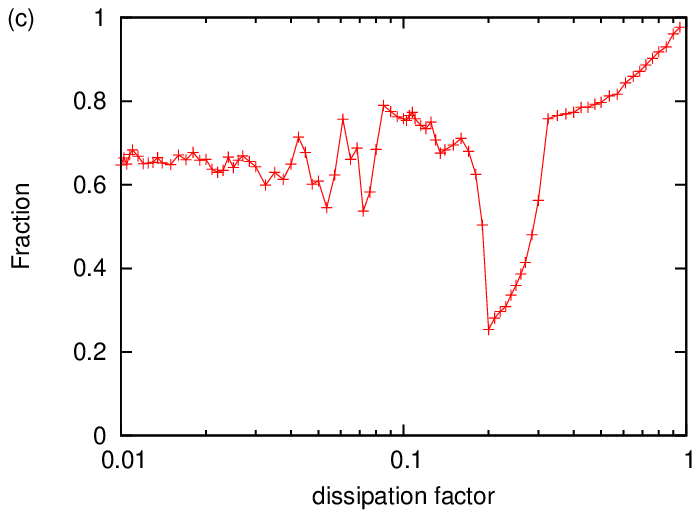} 
\includegraphics[angle=0,width=8.0cm]{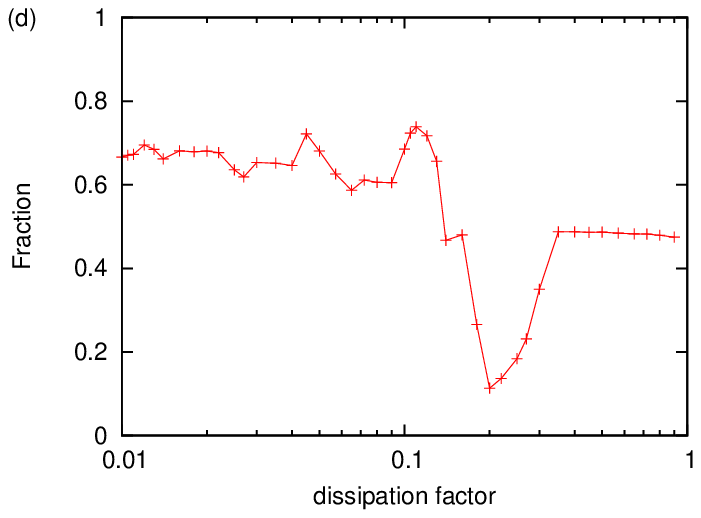} 
\end{center}
\end{figure}

\vspace{0.5in}
 FIGURE 9


\newpage 

\begin{figure}[H]
\begin{center}
\includegraphics[angle=0,width=8.0cm]{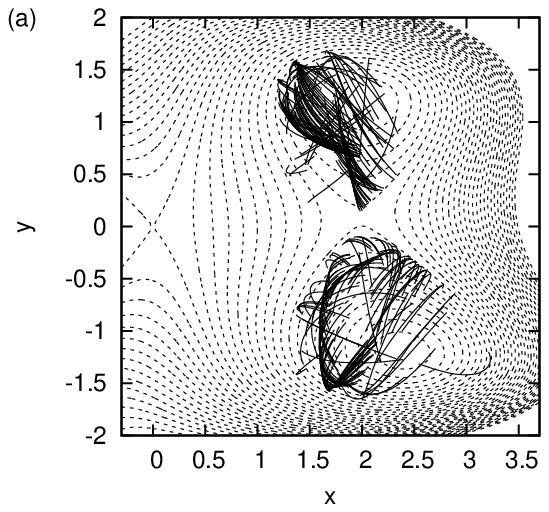} 
\includegraphics[angle=0,width=8.0cm]{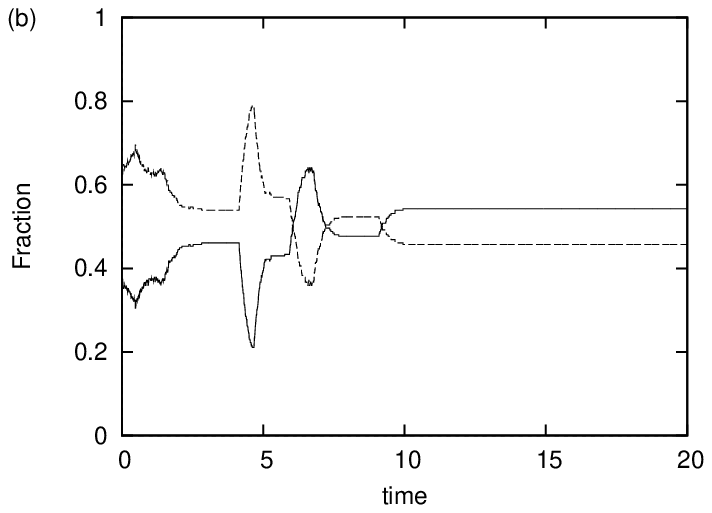} \\
\includegraphics[angle=0,width=8.0cm]{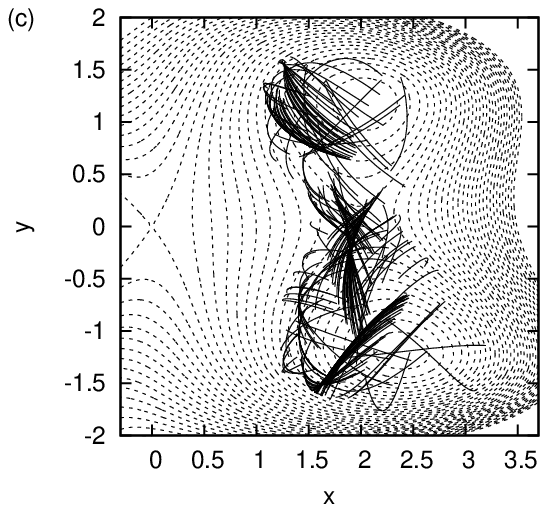} 
\includegraphics[angle=0,width=8.0cm]{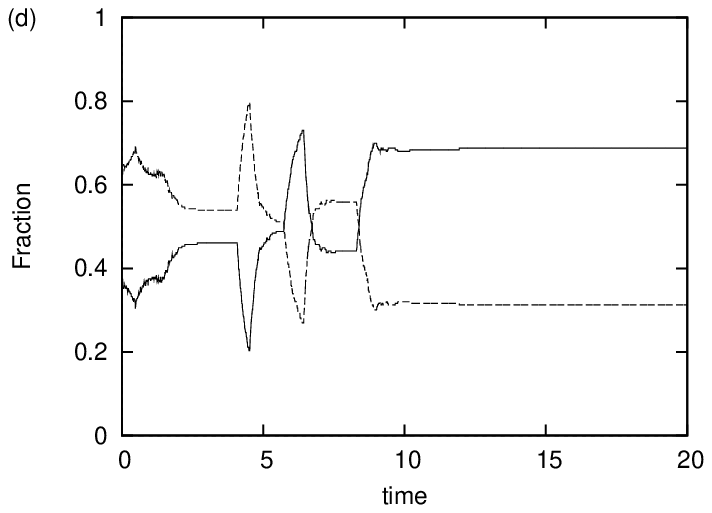} 
\end{center}
\end{figure}

\vspace{0.5in}
 FIGURE 10

\end{document}